\documentclass[aps,preprint]{revtex4}

\usepackage{epsfig}

\begin{document}

\title{Stochastic lattice gas model describing the dynamics of an epidemic}

\author{David R. de Souza and T\^{a}nia Tom\'{e}}

\affiliation{Instituto de F\'{\i}sica, Universidade de S\~{a}o Paulo, \\
Caixa Postal 66318\\
05314-970 S\~{a}o Paulo, S\~{a}o Paulo, Brazil}
\date{\today}

\begin{abstract}

We study a stochastic process describing
the onset of spreading dynamics of an epidemic
in a population composed by individuals of three classes: susceptible
(S), infected (I), and recovered (R).
The stochastic process is defined by local rules and involves
the following cyclic process: S$\to$I$\to$R$\to$S (SIRS).
The open process S$\to$I$\to$R (SIR) is studied as a particular
case of the SIRS process.
The epidemic process is analyzed at different levels of description:
by a stochastic lattice gas model and by a birth and death process.
By means of Monte Carlo simulations and dynamical mean-field
approximations we show that the SIRS stochastic lattice gas model
exhibit a line of critical points separating two phases: an absorbing
phase where the lattice is completely full of S individuals
and an active phase where S, I and R individuals coexist,
which may or may not present population cycles.
The critical line, that corresponds to the onset of
epidemic spreading, is shown to belong in the directed percolation
universality class. By considering the birth and death
process we analyse the role of noise in stabilizing
the oscillations.


\end{abstract}

\maketitle

\section{Introduction}

Among the several theoretical descriptions of
population biology systems, the stochastic approach is 
one capable of revealing the ubiquitous fluctuations observed
in these systems \cite{nisbet,renshaw,keel99,nisbet00,keel02}.
Using a stochastic lattice gas model, 
we study here the dynamics of propagation 
of an epidemic in a population in which the individuals are
separated into three classes
determined by their relative states to a given disease: 
susceptible (S), infected (I) and recovered (R) individuals. 
One of the most well known models in
this context is the so-called susceptible-infected-recovered (SIR) model 
\cite{nisbet,renshaw,keel99,nisbet00,keel02,
ker,hastings,grass,keel,ara,alonso,caliri,aiello},
a model for an epidemic which
occurs during a time interval that is much smaller then the lifetime of the
host. It describes the spreading of an epidemic process occurring
in a population initially composed by
susceptible individuals that become infected by contact with infected
individuals. Once infected the individuals can recover spontaneously
becoming immune.  
An extension of the SIR model, to be considered here, is obtained
by allowing the recovered individual to becoming susceptible.
This extension is called  
susceptible-infected-recovered-susceptible (SIRS) model
and will be the main concern of this paper. The SIR model
will be then a particular case of the SIRS model.

The stochastic dynamics is studied on a two-dimensional lattice that
represents the space where the individuals live. 
Each site of the lattice is occupied by just one individual that
can be in one of the three states: susceptible (S), infected (I) or
recovered (R). The stochastic
rules are such that S becomes I by a catalytic reaction,
I turns into R spontaneously, and R becomes S also spontaneously. 
By means of Monte Carlo simulations, 
dynamic mean-field approximations and
also by means of Fokker-Planck and Langevin equations, we study the dynamics
and the onset of an epidemic spreading. Our results show that the
stationary states 
described by the SIRS stochastic lattice gas model, 
can be of two types: an absorbing (or inactive) state, where the lattice is
fully covered by susceptible, and an active state. In the active state
the processes of infection, recovering and loss of immunity are
continuously occurring.
The phase transitions that take place in 
this model are studied in detail, as well as the critical 
behavior, and critical exponents are obtained from
Monte Carlo simulations. We remark that the transitions
are nonequilibrium ones because the SIRS stochastic lattice gas model
belongs to the class of models that are intrinsically 
irreversible. In other words, 
the dynamics do not obey detailed balance.

From the stochastic lattice gas model we derive a birth and death process
for the numbers of individuals of each class and analyze relevant
quantities to characterize the epidemic spreading such as the time
correlation functions. We find that the SIRS model presents
stochastic oscillations on a certain region of the space of parameters,
inside the active region.
These oscillations are better understood if
the system is described by a Langevin equation from which it becomes
clear that the amplitude of these oscillations decay
as the system size increases and that the noise
enhances the oscillations \cite{alonso,morita,boland,lugo,tome09}.
The Fokker-Planck and the associate
Langevin equations are derived from the birth and death process
which in turn is obtained from the original SIRS stochastic
lattice gas model.  
The derivation of the birth and death process 
is obtained by a contraction procedure in which the
degrees of freedom  is reduced to a few variables \cite{tome09}. 

We have also analyzed the critical behavior of the model
occurring around the transition from the active to the nonactive
state. We find that the critical behavior places the SIRS
but not the SIR model into the directed percolation (DirP) 
universality class. 
The critical behavior is obtained by determining 
the static and dynamic critical exponents by means of 
numerical simulations.
The SIR stochastic lattice gas model 
studied here is shown to belong to
a distinct class, namely the dynamic percolation (DynP) 
universality class in agreement with the conjecture
by Grassberger \cite{grass}.

\section{Stochastic lattice gas model}

The SIRS stochastic lattice gas model is defined on a regular lattice
of $N$ sites
in which each site can be occupied by either one susceptible individual
(state S), or one infected individual (state I), 
or one recovered individual (state R). 
The dynamics consists of three subprocesses: 
(1) I $+$ S $\to$ I $+$ I, auto-catalytically; 
(2) I $\to$ R, spontaneously; and
(3) R $\to$ S, spontaneously. 
At each time step a site is randomly chosen. (i) If the site is 
in the state S and there is at least one neighboring site in
the state I then S becomes I with probability proportional to 
a parameter $b$ and the number of neighboring I sites;
(ii) if the site is in state I it becomes R spontaneously
with probability $c$; and (iii) if the site is in state R it becomes
S spontaneously with probability $a$. 

At each site $i$ of a square lattice we
attach a variable $\eta _{i}$ that takes the values 0, 1 or 2, 
according to whether the site is in the state R, S, or I, 
respectively. The allowed transitions
of the states of a site are cyclic, that is, $0\to 1\to
2\to 0$ and the corresponding transition rate
is given by 
\begin{equation}
w_i(\eta) = \frac{b}4 \delta(\eta_i,1)
\sum_j \delta(\eta_j,2) + c\delta(\eta_i,2)+ a\delta(\eta_i,0),
\label{1}
\end{equation}
 where $\eta =(\eta_1,\ldots,\eta_i,\ldots,\eta_N)$ 
denotes the whole microscopic configuration, 
$\delta(x,y)$ denotes the Kronecker delta,
the summation is over the nearest neighbours of site $i$ and $b$, $c$
and $a$ are parameters related to the subprocesses (i), (ii) and (iii)
described above. 
The transition rates of the SIR model is regarded as a particular
case of the SIRS transition rate given by equation (\ref{1}), 
obtained by setting $a=0$, which means that the 
subprocess (3) described above is suppressed. 

The time evolution of the probability distribution $P(\eta,t)$ at time $t$
obey the master equation 
\begin{equation}
\frac{d}{dt}P(\eta,t)=\sum_i\{w_i(\eta^i)P(\eta^i,t)-w_i(\eta)P(\eta,t)\},
\label{2}
\end{equation}
where the configuration $\eta^{i}$ is obtained from configuration
$\eta$ by an anticyclic permutation of the state of site $i$, that is, 
$2\to1\to 0\to 1$.

By rescaling time in the master equation  
it is always possible to reduce
the three parameter of the model to just two. Thus
we choose the parameters in such a way that $a+b+c=1$
so that $a$, $b$ and $c$ can be interpreted as probabilities;
this allows us to introduce a new parameter $p$ such that
$a$ and $b$ are given by
\begin{equation}
a=\frac{1-c}2-p \qquad{\rm and}\qquad b=\frac{1-c}2+p,
\end{equation}
where $-1/2\leq p \leq 1/2$ and $0\leq c\leq 1$.


\begin{figure}[tbp]
\centering
\epsfig{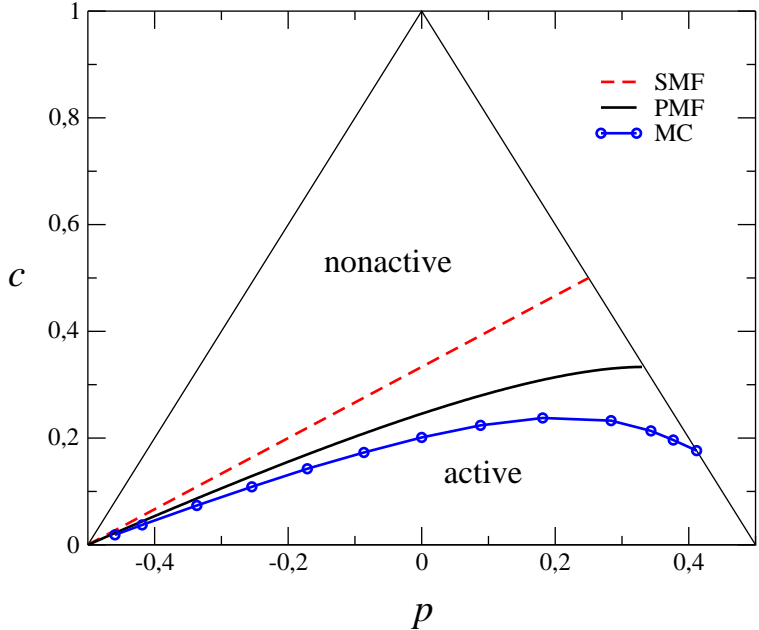}
\caption{(Color online) Phase diagram of the SIRS stochastic lattice gas 
model in the p-c space of parameters showing the two phases: a nonactive
phase, with only susceptible individuals, 
and an active phase, with the three types of individuals,
where the disease spread over the population. 
The phase boundaries were obtained by Monte Carlo (MC) simulations,
by pair mean-field (PMF) and simple mean-field (SMF) approximations.
The right edge of the triangle, where $a=0$,
corresponds to the SIR model.}
\label{diag}
\end{figure}

\begin{table}
\caption{Critical exponents from Monte Carlo simulations 
for the SIRS model and SIR model ($a=0$) defined on 
regular square lattice. For comparison we show the 
exponents related to the directed percolation (DirP)
universality class
and to the dynamic percolation (DynP) universality class
according to reference \cite{dick}.}
\label{table1}
\begin{center}
\begin{tabular}{lllllll}
\hline
\hline
 $a$ & $c$    & $\beta$ & $\beta/\nu$ & $\eta$ & $z$  & $\delta$ \\ 
\hline
\hline
0.00 & 0.1765(5)  & 0.129(10) & 0.096(7) & 0,586(1) & 1,773(1) & 0,094(3) \\ 
\hline
DynP & & 0.1389    & 0.1042   & 0.586    & 1.771    & 0.092    \\
\hline
\hline
0.025& 0.19640(5) & 0.574(13) & 0.806(11)& 0.22(2)  & 1.2(1)   & 0.42(3)  \\ 
0.05 & 0.21355(5) & 0.56(2)   & 0.805(14)& 0.232(7) & 1.163(7) & 0.424(15)\\ 
0.10 & 0.23270(5) & 0.56(2)   & 0.83(3)  & 0.230(8) & 1.126(10)& 0.435(12)\\ 
0.20 & 0.2377(1)  & 0.56(2)   & 0.81(2)  & 0.227(2) & 1.132(2) & 0.446(2) \\
\hline
DirP & & 0.583(4)& 0.795(10)& 0.2295(10)& 1.1325(10)&0.4505(10)\\ 
\hline
\hline
\end{tabular}
\end{center}
\end{table}

We have obtained the phase diagram of the SIRS
stochastic lattice gas model defined on a regular square lattice
by using Monte Carlo simulations.
The phase diagram in the space of parameters $p$ and $c$
is shown in figure \ref{diag}. 
The model exhibits
an absorbing susceptible state and an active state where individuals are
continuously changing their states. We have located the critical line
separating the two phases and determined the dynamic and static
critical exponents. Both are in agreement with the ones of the direct
percolation universality class. The dynamic exponents were 
obtained by time dependent Monte Carlo simulations \cite{marro} and the 
static exponents by Monte Carlo simulations performed on finite square
lattices with a number of sites up to 256$\times$256.
In the last case we have used a finite size scaling 
proposed in reference \cite{ttmj05}. 
Our best results for the exponents are given in table \ref{table1}. 
Our estimates for the SIRS model ($a>0$) are in agreement 
with the ones for the directed percolation universality class \cite{dick}.

For the SIR stochastic lattice gas model ($a=0$) we have used the 
same Monte Carlo
procedures to determine the static and dynamic critical exponents.
Our estimates of the exponents, shown in table \ref{table1}, 
are in agreement with the ones related to the dynamic percolation 
universality class \cite{grass,dick}.
We remark that the active phase 
of the SIR stochastic lattice gas model
is characterized by infinitely many absorbing configurations.
These absorbing configurations are reached when the number of 
infected individuals disappears.

It is worth mentioning that when the values of the parameters $b$ and $c$ 
are much smaller than $a$ and each other of the same order, corresponding
to the region around the left corner of the triangular phase diagram
of figure \ref{diag}, an R individual is almost instantaneously 
converted into an S individual. We may therefore replace
the two spontaneous processes I$\to$R and R$\to$S by just one
spontaneous process I$\to$S. The SIRS model is thus reduced to the 
process S$\to$I$\to$S, composed by a catalytic infection and
a spontaneous recovering, defining a class of models called SIS
\cite{keel,martins}; 
in the present case it corresponds to the contact process,
known to be in the DP universality class \cite{marro}. In the limit
of small values of $b$ and $c$, the ratio $b/c$ is identified
as creation rate $\lambda$ of contact process, so that the transition
line around the left corner of the triangular phase diagram
is given by $b/c=\lambda_c$ where $\lambda_c$
is the critical infection rate of the contact process defined
on a regular square lattice. 

\section{Birth and death process}

\begin{figure}[tbp]
\centering
\epsfig{file=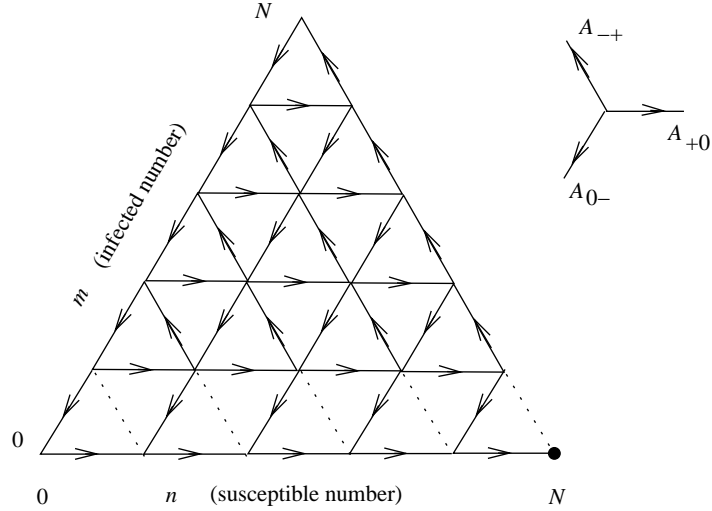,width=9.5cm}
\caption{Transitions of the birth and death stochastic process in the space
of susceptible and infected numbers for the SIRS model. The transition to
the east represents a recovered individual becoming susceptible with rate 
$A_{+0}$, to the northwest a susceptible individual becoming infected with
rate $A_{-+}$, to the southwest an infected individual becoming recovered with
rate $A_{0-}$. The full circle represents an absorbing state. }
\label{bdsirs}
\end{figure}

A distinct stochastic description of the system under discussion 
is the one known as birth and death process or one-step process
\cite{van}.
In this description the state of the system is characterized
by a few stochastic variables. 
A birth and death process is here obtained from the SIRS
stochastic lattice gas model by reducing 
the number of degrees of freedom. 
We use a contraction procedure leading to a description 
with just two stochastic variables: the number $n$
of susceptible individuals and the number $m$ of infected
individuals. These variables are defined by
\begin{equation}
n = \sum_i^N \delta(\eta_i,1)
\qquad{\rm and}\qquad
m = \sum_i^N \delta(\eta_i,2).
\end{equation}
The appropriate stochastic description
involves the probability $P(n,m,t)$ of $n$ and $m$ at time $t$ which is
assumed to obey the following birth and death master equation 
\[
\frac{d}{dt}P(n,m,t)=N\sum_{\sigma =-1}^{1}\sum_{\tau =-1}^{1}\{A_{\sigma
\tau }(n-\sigma ,m-\tau )P(n-\sigma ,m-\tau ,t)
\]
\begin{equation}
-A_{\sigma \tau}(n,m)P(n,m,t)\}.
\label{10}
\end{equation}
This equation comes from the master equation (\ref{2})
by summing over $\eta $ with
the restriction that the number of susceptible individuals is $n$
and the number of infected individuals is $m$.
This procedure leads to the following results 
\begin{equation}
A_{+0}=az,  \label{11}
\end{equation}
\begin{equation}
A_{-+}=bxy,  \label{12}
\end{equation}
\begin{equation}
A_{0-}=cy,  \label{13}
\end{equation}
where $z=1-x-y$, $x=n/N$ and $y=m/N$, and $N$ is the population size. 
Expressions (\ref{11}) and (\ref{13}) are exact
and expression (\ref{12}) is derived by a scheme
similar to the so called simple mean-field approximation
\cite{tome09}.

\begin{figure}[tbp]
\centering
\epsfig{file=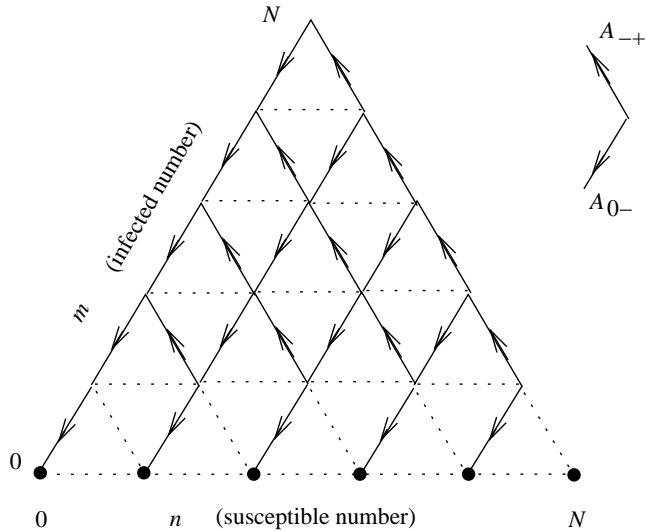,width=8.5cm}
\caption{Transitions of the birth and death stochastic process 
in the space of susceptible and infected numbers for the SIR model. 
The transition to the northwest represents a susceptible individual 
becoming infected with rate $A_{-+}$, to the southwest an infected 
individual becoming recovered with rate $A_{0-}$. A recovered 
individual never becomes susceptible. The full circles
represent absorbing states. }
\label{bdsir}
\end{figure}

The birth and death process defined by the master
equation (\ref{10}) can be regarded
as a random walk in the space $(n,m)$ as shown in Figure \ref{bdsirs}. The
possible jumps are (a) $(n,m)\to(n+1,m)$, with probability $A_{+0}$, 
(b) $(n,m)\to(n-1,m+1)$, with probability $A_{-+}$, and 
(c) $(n,m)\to(n,m-1)$ with probability $A_{0-}$.

\begin{figure}[tbp]
\centering
\epsfig{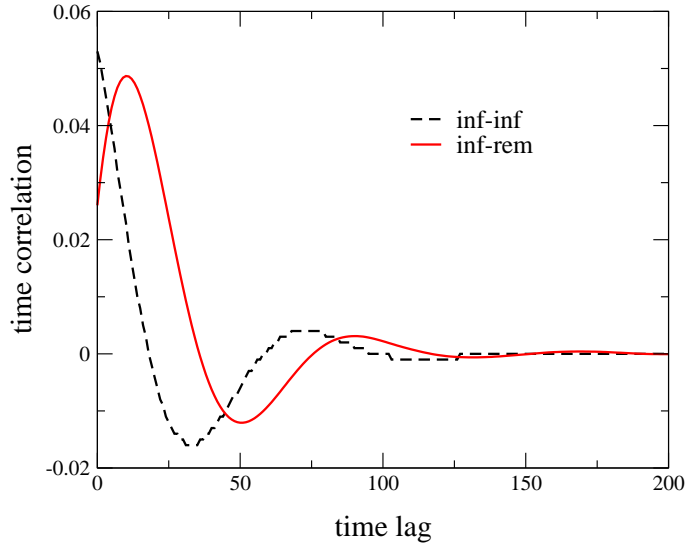}
\caption{(Color online) Infected-infected (inf-inf) time autocorrelation
function and infected-removed (inf-rem) time cross-correlation function
versus the time lag for $p=0.25$ and $c=0.40$. Results from numerical
simulation of the birth and death process.}
\label{correl}
\end{figure}

We have simulated the birth and death stochastic process 
according to the transition rates given by equations (\ref{11}), (\ref{12})
and (\ref{13}) for $N=5000$. We have also determined the time correlation
functions as a function of the time lag $t$ defined by
\begin{equation}
c_{fg}(t) = \int [f(t^\prime+t)-\bar f][g(t^\prime)-\bar g]dt^\prime,
\end{equation}
where $f$ and $g$ can be any of the variables 
$n$ or $m$. As an example,
we show in figure \ref{correl} the auto-correlation function
for infected individuals and the cross-correlation between
infected and removed individuals for the following values
of the parameters: $p=0.25$ and $c=0.40$.
The damped oscillations of the time correlation function
indicates that the populations of individuals in each class
oscillate in time with period of $T=72(3)$ Monte Carlo steps.
This period was confirmed by Fourier analysis
of the time correlation functions.
The active phase can thus be separated into two
regions as can be seen in figure \ref{linha}.
One region corresponding to sustained oscillations
in population of individuals and the other without
oscillations. 

\section{Langevin equations}

Considering the situation where the population size $N$ is large 
we perform an expansion of the master equation (\ref{10}) 
in powers of $1/N$ \cite{van,gar}. Up to second order 
in $1/N$ we get a Fokker-Planck equation for the 
probability density ${\cal P}(x,y,t)$,
namely
\[
\frac{\partial}{\partial t}{\cal P} = 
-\frac{\partial}{\partial x}(f_1{\cal P})
-\frac{\partial}{\partial y}(f_2{\cal P})
\]
\begin{equation}
+\frac{1}{2N}\left(\frac{\partial^2}{\partial x^2}(D_{11}{\cal P})
+2\frac{\partial^2}{\partial x\partial y}(D_{12}{\cal P})
+\frac{\partial^2}{\partial y^2}(D_{22}{\cal P})\right),
\label{fp}
\end{equation}
where 
\begin{equation}
f_{1}(x,y) = az-bxy,
\qquad \qquad 
f_{2}(x,y) = bxy-cy,
\end{equation}
and 
\begin{equation}
D_{11}(x,y) = az+bxy,
\end{equation}
\begin{equation}
D_{12}(x,y) = -bxy,
\end{equation}
\begin{equation}
D_{22}(x,y) = bxy+cy.
\end{equation}
The Langevin equations associated to the Fokker-Planck 
equation (\ref{fp}) are
\begin{equation}
\frac{dx}{dt} = f_{1}(x,y)+\frac{1}{\sqrt{N}}\,\zeta _{1}(t),  
\label{lang1}
\end{equation}
\begin{equation}
\frac{dy}{dt} = f_{2}(x,y)+\frac{1}{\sqrt{N}}\,\zeta _{2}(t),  
\label{lang2}
\end{equation}
where $\zeta_{i}(t)$, with $i=1,2$, are white Gaussian noise functions with
zero mean obeying the properties 
\begin{equation}
\langle \zeta _{i}(t)\zeta_{j}(t^{\prime })\rangle =D_{ij}(x,y)\,\delta
(t-t^{\prime }).
\end{equation}
Similar expansion procedure to study population biology systems
has been used by other authors 
\cite{nisbet,renshaw,morita,boland,lugo,tome09}.

\section{Mean-field approximation}

\begin{figure}[tbp]
\centering
\epsfig{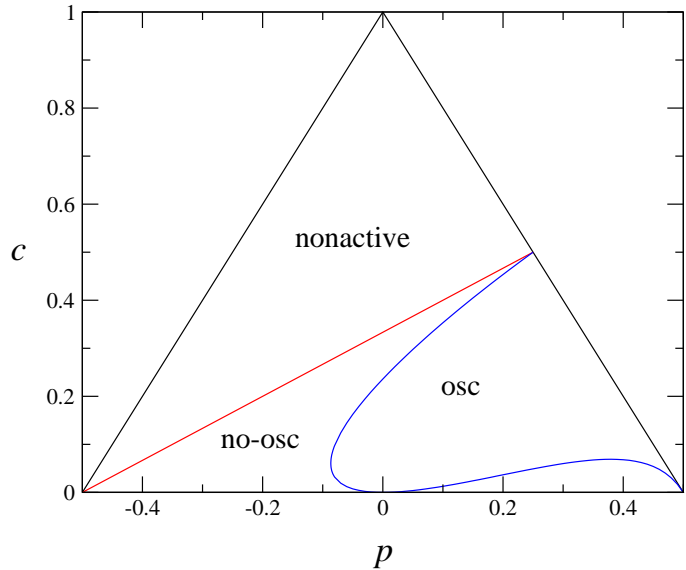}
\caption{(Color online) Phase diagram of the SIRS model according to
the birth and death process. Inside the active phase,
below the nonactive phase,
there are two regions corresponding to the oscillatory (osc) and 
to the nonoscillatory (non-osc) regimes. The line of separation
between these two regimes is given by equation (\ref{sep}).}
\label{linha}
\end{figure}

In the limit $N\to\infty$ the noise disappears and we are left with a
deterministic dynamics which is identified here as a dynamic simple 
mean-field approach.
The time evolution of the densities is given by 
\begin{equation}
\frac{dx}{dt} = az-bxy,
\end{equation}
\begin{equation}
\frac{dy}{dt} = bxy - cy.
\end{equation}
The equation for $z$ is not necessary because $x+y+z=1$.
The trivial solution is $(x^*,y^*)=(1,0)$ corresponding to 
a state where the entire population is composed by
susceptible individuals only, called the inactive phase. 
The nontrivial solution is given by
\begin{equation}
x^*=c/b, 
\qquad\qquad 
y^*=a(b-c)/b(a+c),  
\label{29a}
\end{equation}
and is valid as long as $b/c>1$,
corresponding to a state of epidemic spreading,
where the individuals of the three classes coexist.
The stability of the solutions are obtained from the eigenvalues
of the Jacobian matrix
\begin{equation}
J = \left( 
\begin{array}{rr}
-a-by^* & \;\;-a-bx^* \\ 
by^*    & -c+bx^* \\
\end{array}
\right).
\end{equation}
For the trivial solution the eigenvalues are $\,-a$ and $\,b-c$. 
Therefore this solution is stable as long as $b/c<1$. 
The transition from the inactive phase to the active phase
occurs along the line defined by $b=c$ or $c=(1-2p)/3$, 
as shown in figure \ref{diag} and \ref{linha}.
For the nontrivial solution, corresponding to the active state,
the eigenvalues $\lambda$ are given by
\begin{equation}
\lambda = \frac12 \left( -\frac{a(a+b)}{a+c} \pm \sqrt{\Delta} \right),
\end{equation}
where 
\begin{equation}
\Delta = a^2 \left(\frac{a+b}{a+c} \right)^2 - 4a(b-c).
\end{equation}
Therefore,
when $\Delta<0$, the eigenvalues are complex and 
the solution approaches an asymptotic stable focus.
When $\Delta>0$, they are real and the solution
approaches an asymptotic stable node.
The transition from a focus to a node occurs when $\Delta = 0$, 
that is, when
\begin{equation}
a^2 (a+b)^2 = 4a(b-c)(a+c)^2,
\label{sep}
\end{equation}
which defines the separation line between these two behaviors,
as shown in figure \ref{linha}.

Let us find some particular solutions of the separation 
line (\ref{sep}).
When $a=b$ (or $p=0$) this equation reduces to
$c^2 + 4c -1= 0$ which gives $c= \sqrt{5} - 2 = 0.236068$.
When $b=a+1/2$ (or $p=1/4$), equation (\ref{sep}) gives two solutions:
one is $c=1/2$ and the other is the solution of the root of
$c^2+(5/2)c-(1/8)=0$ which gives
$c = (3\sqrt{3}-5)/4= 0.049038$.
Improvements at qualitative level in the phase diagram can be achieved by
considering dynamic mean-field approximations which takes into account
correlations between neighbors sites. This procedure 
has been largely used in the context
of stochastic lattice gas models, also called 
{\it interacting particle systems},
describing population biology systems
\cite{ara,morita,tome09,durrett,sat94,sat97,aguiar,lebo}.
By using a pair mean-field approximation (PMF) \cite{sat94} we find for the 
present model a transition
line between the active phase and the inactive phase which
is shown in figure \ref{diag}.

\section{Conclusions}

In this article we have examined a spatial-structured model 
describing the dynamic of a SIRS epidemic process.
This model shows a continuous phase transition between a
phase where the disease spread and other where it dies out in a short time.
The phase diagram of the SIRS stochastic lattice gas model was 
obtained by Monte Carlo simulations and dynamic mean-field
approximations.
Our Monte Carlo results indicate that the SIRS 
stochastic lattice gas model exhibits a line of critical points
that belongs in the directed percolation universality class.
As far as we know this is the first determination of the universality 
class of the SIRS stochastic lattice gas model.

We have also derived a birth and death process for the
numbers of individuals in each population class. 
This process was derived
from the stochastic lattice gas model in the scope of simple 
mean-field approximation. A Fokker-Planck equation as well as the 
associate Langevin equations were then
derived from the birth and death process. 
The simulation of this process, 
which can be seem as a random walk
in the space of the number of susceptible and infected individuals,
shows that the active state is subdivided into two regions:
one in which the numbers of individuals in each class oscillate
in time and another where they remain constant in time.

In summary, we have 
devised and analyzed the SIRS epidemic process
at different levels of description: (a) the
spatial, stochastic and local description, given by 
the SIRS stochastic lattice gas model; (b) the
stochastic but zero dimensional birth and death process; and
(c) the deterministic level in which the space is not considered,
given by the simple mean-field approximation where all individuals are
well mixed.
The fundamental level is the one based on stochastic
lattice gas models also called interacting particle systems
\cite{durrett}. These spatial structured models 
are relevant when the diseases spreads over a 
region with few individuals 
or when a group of individual of a given class becomes
isolated, loosing contact with the other class of individuals
during a certain period of time \cite{durrett00,kelly}.

\section*{Acknowledgment}

We acknowledge helpful discussions with M. J. de Oliveira
and financial suppport from the Brazilian
agency CNPq.


\end{document}